\documentclass[conference]{IEEEtran}
\IEEEoverridecommandlockouts
\usepackage{cite}
\usepackage{amsmath,amssymb,amsfonts}
\usepackage{booktabs}
\usepackage{algorithmic}
\usepackage{graphicx}
\usepackage{textcomp}
\usepackage{xcolor}
\def\BibTeX{{\rm B\kern-.05em{\sc i\kern-.025em b}\kern-.08em
    T\kern-.1667em\lower.7ex\hbox{E}\kern-.125emX}}

\usepackage{amsmath}
\usepackage{amssymb}
\usepackage{mathtools}
\usepackage{amsthm}
\usepackage{algorithm}
\usepackage{tikz}
\usetikzlibrary{positioning, fit, arrows.meta, shapes}
\newcommand{\empt}[2]{$#1^{\langle #2 \rangle}$}
\usepackage{pgfplots}
\pgfplotsset{compat=1.16}
\usepackage{blindtext}
\usepackage{hyperref}
\usepackage{longtable}
    
\begin{document}

\title{Deep Learning-Based Speech and Vision Synthesis to Improve Phishing Attack Detection through a Multi-layer Adaptive Framework\\
{\footnotesize \textsuperscript{*} }
\thanks{Identify applicable funding agency here. If none, delete this.}
}

\author{\IEEEauthorblockN{1\textsuperscript{st} Tosin Ige}
\IEEEauthorblockA{\textit{Dept. of Computer Science} \\
\textit{The University of Texas at El Paso}\\
Texas, USA \\
toige@miners.utep.edu}
\and
\IEEEauthorblockN{2\textsuperscript{nd} Christopher Kiekintveld}
\IEEEauthorblockA{\textit{Dept. of Computer Science} \\
\textit{The University of Texas at El Paso}\\
Texas, USA \\
cdkiekintveld@utep.edu}
\and
\IEEEauthorblockN{3\textsuperscript{rd} Aritran Piplai}
\IEEEauthorblockA{\textit{Dept. of Computer Science} \\
\textit{The University of Texas at El Paso}\\
Texas, USA \\
apiplai@utep.edu}
}

\maketitle

\begin{abstract}
The ever-evolving ways attacker continues to improve their phishing techniques to bypass existing state-of-the-art phishing detection methods pose a mountain of challenges to researchers in both industry and academia research due to the inability of current approaches to detect complex phishing attack. Thus, current anti-phishing methods remain vulnerable to complex phishing because of the increasingly sophistication tactics adopted by attacker coupled with the rate at which new tactics are being developed to evade detection. In this research, we proposed an adaptable framework that combines Deep learning and Randon Forest to read images, synthesize speech from deep-fake videos, and natural language processing at various predictions layered to significantly increase the performance of machine learning models for phishing attack detection.
To validate both the effectiveness and adaptability of our proposed framework in overcoming limitations in current approaches and its ability to detect complex phishing site, we created 4 categories of phishing sites and uploaded them to a secure server with a compromised DNS on a friendly URL; the first was a text-only phishing site, image-only phishing site, video-only phishing site, and a phishing site combining all the features. We use SEO friendly URLs, and hacked legitimate DNS on the text-only phishing site, so that they can evade detection at 1\textsuperscript{st} layer until the 4\textsuperscript{th} layer of the framework where they were detected, we also created phishing sites where text are in image only format, text-only, and video only format using deep-fake video to test the adaptability of our proposed framework to different scenarios of a sophisticated or complex phishing site, our proposed framework successfully overcome limitations in existing approaches, significantly improve phishing attack detection,  and successfully detect complex phishing webpages with multi-dimensional deep-fake videos, images, and texts. 
\end{abstract}


\section{Introduction}
The insufficiency of traditional phishing detection methods such as user education \cite{sarker2024multi} and rule-based methods \cite{jain2022survey} against sophisticated phishing attack techniques has led researchers to exploration of possible AI-based solutions. While several machine learning-based models have been proposed, the fact that attackers use advanced innovative methods that are continuously changing to carry out phishing attacks renders previously proposed machine learning models ineffective against sophisticated attacks \cite{ige2023adversarial}. Although tools like PhishTank, and OpenPhish were created for the effective detection of malicious Uniform Resource Locator (URL) the rate at which malicious websites are created coupled with the sophistication of the deception method easily overwhelm the system as phishing sites are being created every 11 seconds \href{https://dataprot.net/statistics/phishing-statistics/} {according to dataprot 2023 phishing statistic report}.

One of the common problems with the current ML model is the quality of the dataset used for the training model which has a significant impact on both the accuracy and overall performance of the model \cite{uccar2020effect}. These data do not reflect the ever-changing strategies through which attackers continue to fool existing machine learning-based models to evade detection. In addition, the balancing problem between human factors and model accuracy causes illegal flagging of newly registered legitimate websites due to weak domain authority. Phishing websites have a short life span as they are quickly taken down before detection and another one is created, It is the rate at which an existing phishing website is taken down after launching a campaign and the immediate creation of a new phishing website \cite{alkhalil2021phishing} to begin another campaign coupled with the ever-changing but sophisticated techniques that makes the problem very potent and significant. 

While several models and machine learning-based frameworks have been proposed, the ever-evolving ways attacker increases the sophistication of phishing attack to bypass existing state-of-the-art anti-phishing detection and prevention systems pose a mountain of challenges leading to the relative ineffectiveness of previously proposed models against a more complex phishing attack. Thus, the constant evolvement and innovation in phishing techniques adopted by attackers are the reason why current detection method remains vulnerable to complex or more sophisticated forms of phishing due to their reliability on \cite{abdulrahman2023web,aljofey2022effective,anitha2022new}, blacklists/whitelists \cite{ghaleb2022phishing}, natural language processing \cite{jain2019machine}, visual similarity \cite{jain2019machine}, rules \cite{jain2016novel},\cite{okomayin2023data},  remains vulnerable to attack due to the following reasons;

\begin{itemize}
\item Having understood how the machine learning-based model works, attackers are now increasingly relying on asymmetrical methods by uploading images and videos to evade detection under various pretexts, and none of the proposed models can single-handedly be effective against such.
\item Very small or minute changes to the uniform Resource Locator (URL) of a blacklisted URL will make the blacklist/ whitelist phishing detection method fail. Also, the fact that there is no worldwide centralized database for whitelisted or blacklisted URLs makes this method even more vulnerable, and so if company X blacklisted my phishing URL on their internal server, I can try it with company Y and be successful.
\item In machine learning phishing detection method that relies on relevant features like URL, webpage content, website traffic, search engine, WHOIS record, and Page Rank have their vulnerabilities because firstly, such classifier will misclassify a phishing URL that is hosted on a hacked or compromised server as benign leading to false negative, secondly using domain age as a feature to train a model will always lead to higher false positive simply because the URL of a newly registered legitimate company website will be misclassified. After all, the domain name was recently registered, the page rank is zero, and with low traffic, and thirdly the fact that parameters for those features are gotten from a third-party website is another concern. What will happen if the third-party website is having a downtime?
\item The issue with the visual similarity-based heuristic method which compares both the pre-stored signature such as images, font styles, page layout, screenshot, and so on of the new website with the old website will have general difficulty in detecting anomalies in a newly hosted phishing site.
\item The fact that the majority of the existing machine learning models are trained based on textual features such as “\#”,”.”, Internet Protocol address, URL Length, domain levels, and so on from the Uniform Resource Locator (URL) does not help as any phisher or attacker with little web technologies can develop what we called "friendly URL” depending on the programming language adopted whether JAVA, C\#, Python, PHP or framework to avoid all those features. With a friendly URL, such models are bound to misclassify leading to an increment in false negative rate.
\end{itemize}

For any Machine learning-based phishing detection method to be effective in real-time combat against phishing attacks, it must address each of the stated reasons above for which existing state-of-the-art anti-phishing methods continue to be vulnerable due to the increasingly sophisticated techniques by which phishing attacks are being carried out. It is worth noting that past research work on phishing attack detection had been largely based on approaches, classification, etc. RASHA ZIENI et al.. \cite{zieni2023phishing} focus their review on list-based, similarity-based, and machine learning-based categories of approaches for phishing detection to identify pending research gap, Angad et al.. \cite{muneer2021survey} focus theirs on the advantages and limitations of existing approaches to phishing detection, while also using discussion of related application scenarios as guidance to propose a new method of anti-phishing detection, Yifei Wang \cite{wang2022survey}  categorizes widely used phishing detection methods into seven categories and summarizes them. All previously proposed models, approaches, and frameworks have common limitation, there limitation was that they are either text-based or URL-based which makes it difficult for them to detect complex phishing attack where the attacker uses  deep-fake videos, deep-fake images, textual-based images, or combination of any with traditional textual content.

In this research, we first reviewed some of the most recent works on phishing detection, and state-of-the-art algorithms from the past 5 years to investigate the performance of state-of-the-art machine learning and deep learning classifiers for phishing detection tasks, before proposing a multi-layered adaptive framework that uses computer vision to read images on a phishing webpage, and condense videos from a webpage to audio before synthesizing the speech into a condensed text to increase detection of a phishing attack. We use a combination of random forest algorithm and Long-Short Term Memory (LSTM) at different layers of the framework for effective coordination. The contributions of our research include the proposal of an adaptive multi-layered framework that uses computer vision to read graphic images, synthesize speech from uploaded videos, and natural language processing at various predictions layered to significantly increase the performance of machine learning-based models for phishing detection. Our artifacts which consist of source code, dataset, images, videos, and audio files for this research had been uploaded to a public GitHub repository for reproducibility of our research. Artifact can be found on GitHub at; \newline 
\textbf{ \href{https://github.com/IGETOSIN1/Deep-Learning-Based-Speech-and-Vision-Synthesis-to-Improve-Phishing-Attack-Detection }{Deep Learning-Based Speech and Vision Synthesis to Improve Phishing Attack Detection through a Multi-layer Adaptive Framework} and also at Code Ocean computational research platform,} with the exception of the internally generated deep-fake video and audio data files for privacy.

\section{Related Work}

\subsection{Natural Language Processing (NLP)}

NLP-based models use existing relationships between sentences, words, or letter parts of a language in a given text dataset. This made us explore the possibility of synthesizing an uploaded video from a phishing webpage to feed our neural network model. NLP architectures use modeling, preprocessing, and feature extraction: 

\textbf{Data preprocessing:} It is imperative for text in a given dataset to be preprocessed into a pattern that the model can easily understand because preprocessing effectively turns every character and word in the dataset into a format that the machine learning classifier can understand to extract useful patterns or learn from them. The fact that algorithms learn from data and the quality of the dataset used in training an ML model directly impacts the performance of that model making AI to be data-centric, and hence, priority is given to data preprocessing during NLP.

NLP stemming and Lemmatization
\begin{equation}
\begin{split}
a &< b \\
ab &< bc \\
abc &> bcd \qquad\qquad \rlap{\text{Lemma~\ref{lem:mylemma}}} \\
abcd &> bcde
\end{split}
\end{equation}

Stemming and lemmatization are the two major data preprocessing tasks for natural language processing. During stemming, there is an end-to-end iteration of each word in the dataset to convert them to their base forms such as the mapping of "university" to "univers", and "calamity" to "calam" while lemmatization uses the word's morphology from vocabulary dictionary to find their corresponding roots.

\begin{equation}
   [T_{i} = \begin{cases}
1       ,   & T \leq 1\\
1+ \beta T, & T > 1 
        \end{cases},
\text{ in which } 
T = \begin{cases}
T_{\mathrm{now}}-T_{\mathrm{last}},   & T_{\mathrm{last}} \neq \mathrm{NULL} \\
T_{\mathrm{now}}-T_{\mathrm{update}}, & T_{\mathrm{last}} = \mathrm{NULL}
    \end{cases}]   
\end{equation}

The final preprocessing stage of NLP is sentence segmentation, this process breaks large text into linguistically meaningful sentences where trivial words such as “an,” “the,” “a,” etc that don't add much meaning or information to the text are removed during stop word removal, next we use tokenization to split every text into words and fragments, the result is a combination of word index and tokenized text which could be represented by a numerical token 
 before feeding them to any of the deep learning or machine learning models for prediction.
\subsection{Long Short-Term Memory (LSTM)}
For this research, We opted for Long Short-Term Memory a variant of recurrent neural network (RNN) because of its effective solution to vanishing and exploding gradients which are Long-term dependency problems in Recurrent Neural Networks. The most important functioning part of an LSTM network is the cell state which serves as a memory to the network thereby enabling it to remember the past. Hence their suitability for capturing long-term dependencies and sequence prediction problems \cite{adewale2023encoder}. LSTM network has an input gate, a forget gate, and an output gate which are sigmoid activation functions with an output value of 0 or 1.

\begin{tikzpicture}
\begin{axis}[
    axis lines=middle,
    xmax=10,
    xmin=-10,
    ymin=-0.05,
    ymax=1.05,
    xlabel={$x$},
    ylabel={$y$}
]
\addplot [domain=-9.5:9.5, samples=100,
          thick, blue] {1/(1+exp(-x)};
\end{axis}
\end{tikzpicture}

 It was easy to use the sigmoid function as a gate because we are only given out positive values that could give a straight answer on whether a particular feature should be kept or discarded.

 \tikzset{elementwiseoperation/.style={circle, draw, inner sep=0pt},
    elementwisefunction/.style={ellipse, draw, inner sep=1pt},
    ct/.style={circle, draw, minimum width=1cm, inner sep=1pt},
    gt/.style={rectangle, draw, minimum width=4mm, minimum height=3mm, inner sep=1pt},
%     filter/.style={circle, draw, minimum width=8mm, inner sep=1pt, 
%   path picture={\draw[thick, rounded corners] 
%   (path picture bounding box.center)--++(65:2mm)--++(0:1mm);
%     \draw[thick, rounded corners] 
%   (path picture bounding box.center)--++(245:2mm)--++(180:1mm);}},
    mylabel/.style={font=\scriptsize\sffamily},}

\begin{tikzpicture}[
    % GLOBAL CFG
    font=\sf \scriptsize,
    >=LaTeX,
    % Styles
    cell/.style={% For the main box
        rectangle, 
        rounded corners=5mm, 
        draw,
        very thick,
        },
    operator/.style={%For operators like +  and  x
        circle,
        draw,
        inner sep=-0.5pt,
        minimum height =.2cm,
        },
    function/.style={%For functions
        ellipse,
        draw,
        inner sep=1pt
        },
    ct/.style={% For external inputs and outputs
        circle,
        draw,
        line width = .75pt,
        minimum width=1cm,
        inner sep=1pt,
        },
    gt/.style={% For internal inputs
        rectangle,
        draw,
        minimum width=4mm,
        minimum height=3mm,
        inner sep=1pt
        },
    mylabel/.style={% something new that I have learned
        font=\scriptsize\sffamily
        },
    ArrowC1/.style={% Arrows with rounded corners
        rounded corners=.25cm,
        thick,
        },
    ArrowC2/.style={% Arrows with big rounded corners
        rounded corners=.5cm,
        thick,
        },
    ]

%Start drawing the thing...    
    % Draw the cell: 
    \node [cell, minimum height =4cm, minimum width=6cm] at (0,0){} ;

    % Draw inputs named ibox#
    \node [gt] (ibox1) at (-2,-0.75) {$\sigma$};
    \node [gt] (ibox2) at (-1.5,-0.75) {$\sigma$};
    \node [gt, minimum width=1cm] (ibox3) at (-0.5,-0.75) {Tanh};
    \node [gt] (ibox4) at (0.5,-0.75) {$\sigma$};

   % Draw opérators   named mux# , add# and func#
    \node [operator] (mux1) at (-2,1.5) {$\times$};
    \node [operator] (add1) at (-0.5,1.5) {+};
    \node [operator] (mux2) at (-0.5,0) {$\times$};
    \node [operator] (mux3) at (1.5,0) {$\times$};
    \node [function] (func1) at (1.5,0.75) {Tanh};

    % Draw External inputs? named as basis c,h,x
    \node[ct, label={[mylabel]Cell}] (c) at (-4,1.5) {\empt{c}{t-1}};
    \node[ct, label={[mylabel]Hidden Layer 1}] (h) at (-4,-1.5) {\empt{h}{t-1}};
    \node[ct, label={[mylabel]left:Input}] (x) at (-2.5,-3) {\empt{x}{t}};

    % Draw External outputs? named as basis c2,h2,x2
    \node[ct, label={[mylabel]Cell}] (c2) at (4,1.5) {\empt{c}{t}};
    \node[ct, label={[mylabel]Hidden Layer 3}] (h2) at (4,-1.5) {\empt{h}{t}};
    \node[ct, label={[mylabel]left:Hidden Layer 2}] (x2) at (2.5,3) {\empt{h}{t}};

% Start connecting all.
    %Intersections and displacements are used. 
    % Drawing arrows    
    \draw [ArrowC1] (c) -- (mux1) -- (add1) -- (c2);

    % Inputs
    \draw [ArrowC2] (h) -| (ibox4);
    \draw [ArrowC1] (h -| ibox1)++(-0.5,0) -| (ibox1); 
    \draw [ArrowC1] (h -| ibox2)++(-0.5,0) -| (ibox2);
    \draw [ArrowC1] (h -| ibox3)++(-0.5,0) -| (ibox3);
    \draw [ArrowC1] (x) -- (x |- h)-| (ibox3);

    % Internal
    \draw [->, ArrowC2] (ibox1) -- (mux1);
    \draw [->, ArrowC2] (ibox2) |- (mux2);
    \draw [->, ArrowC2] (ibox3) -- (mux2);
    \draw [->, ArrowC2] (ibox4) |- (mux3);
    \draw [->, ArrowC2] (mux2) -- (add1);
    \draw [->, ArrowC1] (add1 -| func1)++(-0.5,0) -| (func1);
    \draw [->, ArrowC2] (func1) -- (mux3);

    %Outputs
    \draw [-, ArrowC2] (mux3) |- (h2);
    \draw (c2 -| x2) ++(0,-0.1) coordinate (i1);
    \draw [-, ArrowC2] (h2 -| x2)++(-0.5,0) -| (i1);
    \draw [-, ArrowC2] (i1)++(0,0.2) -- (x2);

\end{tikzpicture}

In an LSTM network, the Input Gate tells what new information is to be stored in the cell state, the forget gate gives clear instructions by telling what information is to be thrown away from the cell state, while the output gate gives activation to the output for more accurate prediction. It is during this activation which occurs after filtering the cell state that the output goes through the activation function where the output portion to be predicted is determined, and this occurs when the current LSTM block goes through softmax layer to predict value for the current block.

To mitigate the effect of phishing attack, several methods, frameworks had been proposed for phishing attack detection but with varying results, these methods are classified based on their different approaches  which we classified as Non-Machine Learning, machine learning (Bayesian-based, non-Bayesian-based) and deep learning-based. As attackers continue to navigate potential vulnerabilities to existing phishing detection solution, they are begining to rely on several images and uploaded videos rather than traditional text to enable them to evade detection, the inability of existing machine learning-based model to detect such phishing site is a peculiar limitation to existing AI-based solution. Palla Yaswanth and V. Nagaraju \cite{yaswanth2023prediction}  used Huang and Premaratne data from Kaggle repository with an equal number of phishing and legitimate datasets for novel network of phishing predictions with an accuracy of 95\% for naive Bayes and 94.67\% for random forest based on parameter turning. During the comparison of the performances of naive Bayes \cite{ige2023performance} and random forest for detection of phishing sites in a network, there was no testing of the model against sophisticated form of phishing attack and causes of the 5\% failure rate of naive bayes in the research.

Abdul Karim et al.. \cite{karim2023phishing} proposed a hybrid model which combines logistic regression, support vector machine, and decision tree in conjunction with soft and hard voting, the proposed hybrid model used Grid Search Hyper-parameter Optimization, cross fold validation, and canopy feature selection method to select relevant features from the dataset. The proposed hybrid model resulted in an accuracy of 98.2\% by using the only attribute properties of the uniform resource locator. The sole reliance on the attribute of the URL  makes this approach extremely vulnerable to URL manipulations as any attacker with little experience in web technology can use a malicious webpage with a friendly URL to fool the model.

Ishwarya et al. \cite{ishwarya2023seperation} proposed a phishing detection method comprising of Naive Bayes algorithm, SVM, KNN, and random forest including evaluation of the performances of each of the four (4) classifiers in detection of phishing email. The implementation of each classifiers resulted in the highest accuracy of 98.2\% for naive Bayes, albeit the use an imbalance dataset comprising 87\% ham and 13\% spam for the research surely indicate biased in the proposed model, and the problem of Bayesian poisoning was not addressed in the proposed model.

Kamal Omari \cite{Omari2023} used the UCI phishing domains dataset to proposed machine learning-based model for the purpose of investigating Logistic Regression (LR), k-Nearest Neighbors (KNN), Support Vector Machine (SVM), Naive Bayes (NB), Decision Tree (DT), Random Forest (RF), and Gradient Boosting for phishing detection task. Hence, we believed that the 98.1\% accuracy for phishing detection task obtained from the Naive Bayes classifier by Ishwarya et al. (2023) \cite{ishwarya2023seperation} was due to a massive imbalance in the dataset having 87\% ham and 13\% spam which was not addressed, also the proposed model doesn't address detection evasion through uploaded video and images on a phishing webpage.

Ann Zeky et al. \cite{magdacy2023detect} proposed an extraction-based Naive Bayes model for phishing detection with emphasis on the extraction of relevant features like unusual characters, spelling mistakes, domain names, and URL analysis from unseen web pages for effective classification of a website into malicious and benign. By training the proposed model with a relatively balanced dataset of 7000 records in which 54\% are malicious and 46\% are benign leading to an accuracy of 99.1\%. By using a combination of content extraction and URL analysis, we believed the proposed model would not be vulnerable to malicious URLs in the sense that even if the attacker tried to use a friendly URL to deceive the model, that the model does not rely on the properties of URL alone but also uses background webpage extraction means the proposed model will still be able to classify webpages correctly, albeit an attacker will still be able to use Bayesian poisoning.

Nishitha et al. \cite{nishitha2023phishing} compared performances of machine learning algorithms and deep learning for phishing detection classification by implementing KNN, Decision tree, Random Forest, Logistic Regression as machine learning algorithm, convolusional neural network and recurrent neural network as deep learning in which logistic regression and CNN had the best performances with an accuracy of 95\% and 96\% respectively, albeit the proposed model only uses the URL properties and so couldn't be used for a sophisticated phishing attack that relies on images and video content.

Twana and Murat \cite{mustafa2023feature}  while assuming the absence of a single solution to detect most phishing attacks and to investigate the impact of feature selection on Naive Bayes model. They \cite{mustafa2023feature} developed 6 Naive Bayes-based models in which each model involves a single feature selection technique chosen from individual FS, forward FS, Backward FS, Plus-I takeaway-r FS, AR1, and All. The experiment resulted in the Naive Bayes model with  Plus-I takeaway-r feature selection having the best performance with an accuracy of 93.39\% while the Naive Bayes classifier with individual feature selection technique has the least performance with an accuracy of 92.05\% thereby leading to the conclusion that feature selection has a direct impact on the accuracy of phishing detection.

Jaya T et al. \cite{jaya2023appropriate} explored the prospect of using unsupervised learning to cluster spam and ham messages in mail using frequency weight-age of words in the message content in more of a natural language processing task and comparing the performances of each of Random Forest, Logistic, Random Tree, Bayes Net, and Naive Bayes algorithms with LTSM Algorithms for phishing detection. The experiment resulted in LSTM which is deep learning based having an encouraging performance, followed by random forest.

One limitation that is peculiar to each and every previously proposed models, frameworks, and approaches is that they can only detect text-based and URL-based phishing webpages and URLs as they are only trained based on text and properties of the Uniform Resource Locator. Current machine learning and deep learning models are not trained to detect more complex and increasingly sophisticated phishing attack which relies heavily on SEO friendly URL, putting text-on images, and Deep-fake AI generated video to evade detection. Hence, there vulnerabilities to complex form of phishing attack.

\section{Experimental Setup}
\subsection{Dataset}
The complexity of the research means we cannot rely on a single data. So, we use two publicly available datasets. There is no publicly available dataset for video-based, audio-based, and image-based phishing dataset, so we use simulation to internally generate them.

We use the "B" version of Mendeley phishing dataset which was designed as a benchmark dataset for training a machine learning models for phishing detection. It includes 11430 URLs and 87 extracted features from which models could be trained. Features in the dataset are classified further into three (3) different categories in which 56 extracted features are from the structure and syntax URL, 24 features were extracted from the content of the URL correspondent pages, and the remaining 7 features which are features with the greatest impact on prediction outcome are extracted from external services. 50\% of the dataset used are "phishing" while the remaining 50\% are from "legitimate" URLs. This balance of the dataset ensures that the prediction result is not unfairly tilted toward or against a particular category.

The second public dataset that we used was the Spam Message Classification dataset from KAGGLE containing 5157 unique records. The remaining datasets in the form of deep-fake videos and images were simulated and internally generated due to unavailability of such datasets in the public repository.

\subsection{Settings}
The proposed framework has 4 predictive layers, with each layer suitable for 
 a specialized category of dataset to ensure adaptability. We show the results in several settings. \newline\newline
\textbf{• Layer 1 (URL-Based Training):} We did traditional machine learning training on the first layer using the Mendeley phishing dataset. Out of possible 87 features, we use chi2 from sklearn feature selection library to select the best 19 features, having set the hyper-parameter k-value to 7 for optimal result which gave us a combination of the best 19 features. The dataset was split into two such that 80\% was used for training, while the remaining 20\% was used for validation tests. We choose random forest because of its suitability for URL-based phishing detection relative to other classifiers \cite{alazaidah2024website},\cite{shukla2024http},\cite{van2024applicability},\cite{olukoyaheterogeneous},\cite{shaukat2023hybrid},\cite{joshi2023machine}.During iteration, we set both the depth and random variable to several values for optimal result but only observed a small but negligible change in the variation of the accuracy until 39. with depth > 39, the accuracy remains constant, at least till when we increase the randomness of the tree to 1 before observing little change. We finally settled on setting the randomness state to 0 so that each tree remains the same each time it is generated.

\begin{table}[t]
\caption{Negligible impact of max\textunderscore depth and random\textunderscore state on accuracy.}
\label{sample-table}
\vskip 0.15in
\begin{center}
\begin{small}
\begin{sc}
\begin{tabular}{lcccr}
\toprule

%max_depth & random_state & accuracy  \\
max\_depth& random\_state& accuracy  \\

\midrule
30    & 0& 90.1 \\
20 & 0 & 90.0\\
10    & 0 & 88.1\\
40    & 1 & 88.2 \\
50     & 1 & 89.8\\
60      & 1 & 88.1\\
\bottomrule
\end{tabular}
\end{sc}
\end{small}
\end{center}
\vskip -0.1in
\end{table}

\textbf{• Layer 2 (Image Processing):} This is the layer where the Hypertext Markup Language (HTML) of the actual phishing webpage is secretly web-scrapped behind the scenes without any actual navigation for security purposes. The behind-the-scenes mode of web scrapping the HTML content protects the server and the network from potential drive-by attacks that might originate from the phishing site. All the syntax of HTML mark-up language was removed From the extracted HTML by REGEX as we needed only the content within the opening and closing of the body tag which is the section being served by web server to potential victims while on a phishing website, this step securely brings whatever content (textual, videos, or images) that will be served to potential victim into the framework for series of processing, and this effectively ensure that they cannot evade detection.

Next, we wrote an algorithm to iterate through every filtered word in the sentence, returning only the list of words with any image extension. The fact that the webpage was webs crapped means our program automatically returns a list of the full path of those images from the web server where they are hosted. The returned list is further iterated and passed through an Optical Character Recognition (OCR) library which uses computer vision to read content of each images into a raw text message and forward it to the next layer for further processing.

\begin{algorithm}
    \caption{LSTM Model Training for Natural Language Processing (NLP) task}\label{your_label}
    \begin{algorithmic}
        \STATE  $\mathrm{train\_LSTM} (f_i,w_i,o_j)$
        \FOR{epochs = $1$ to $N$}
            \WHILE{$(j\le m)$}
                \STATE Randomly initialize $w_i=\{w_1,w_2,\dots,w_n\}$
                \STATE input $o_j=\{o_1,o_2,\dots,o_m\}$ in the input layer
                \STATE forward propagate $(f_i\cdot w_i)$ through layers until getting the predicted result $y$
                \STATE compute $e=y-y^2$
                \STATE back propagate $e$ from right to left through layers
                \STATE update $w_i$
            \ENDWHILE
        \ENDFOR
    \end{algorithmic}
\end{algorithm}

\textbf{• Layer 3 (Speech Synthesis):} Having successfully web-scrapped the hypertext mark-up language of the potential phishing site behind the scenes, and without any actual navigation for security purposes at the previous layer. Returned content from the previous layer 2 is further iterated through with  "for" loop. "for" loop iterates through every filtered word in the sentence, returning only the list of words with any video extension, the return list is automatically the full path of those videos from the server where they are hosted.

\begin{figure}[ht]
\vskip 0.2in
\begin{center}
\centerline{\includegraphics[width=\columnwidth]{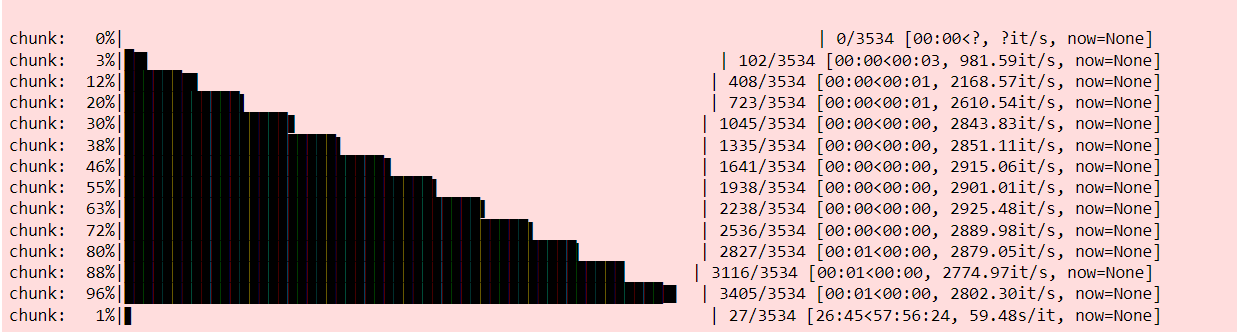}}
\caption{Step-wise speech synthesis of each audio file during execution of "for" loop in layer 3 to produce text which was later passed on to layer 4. Texts from the phishing sites were processed at Layer 1, images were processed at Layer 2, while Layer 3 processed videos. All text was finally outputted to layer 4 for final prediction using a variant of Recurrent Neural Network in Long Short-Term Memory.}
\label{icml-historical}
\end{center}
\vskip -0.2in
\end{figure}

Next, we did further iteration through each of the returned video files, and on each iteration step, we used a combination of gtts, pydud, and moviepy for conversion from video file to audio file ".wav" format, after which the actual synthesis of each speech across the "loop" began with natural language processing speech recognition. The final operation output at this layer is a raw text file obtained from synthesizing the speech. At this stage, we have the images read to text from the previous layer 2 and speeches in the video synthesized to text, next, we combined each of the text from layer 1, layer 2, to layer 3 forwarding them to layer 4 for final prediction. \newline\newline
\textbf{• Layer 4 (Speech Synthesis):} We choose LSTM network because of the effective solution it offers to vanishing and exploding gradient which are Long term dependency problem in Recurrent Neural Network, the cell state in LSTM network serves as a memory to the network thereby given it the ability to remember the past. At layer 4, we have all outputted and processed text contents from each of the previous layers, and there is need to capture every long-term dependencies, short term dependencies, and sequences which could be provided by the cell state in LSTM network to ensure a more accurate prediction.

We built an LSTM deep learning-based model, in which 80\% of 5572 samples were used as training samples while the remaining 20\% was used for validation. The dataset has a maximum of 10,000 features from the word sample, out of which we have 9004 unique words from the dataset. During training and validation, we had wide validation loss leading to low prediction ability but continued to adjust the number of layer, features, epoch, batch size, and activation. We obtained the best result at the following parameter; \newline 
dense layer = 1\newline 
activation layer= sigmoid \newline 
epochs = 10 \newline 
batch size=60 \newline 
feature size = 32 
\begin{figure}[ht]
\vskip 0.2in
\begin{center}
\centerline{\includegraphics[width=\columnwidth]{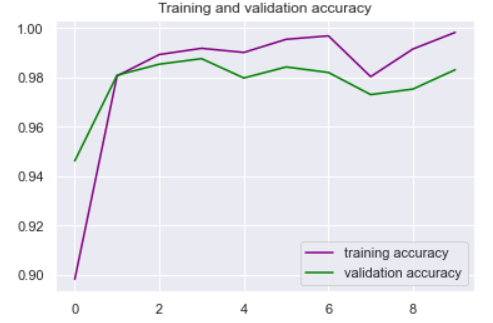}}
\caption{LSTM network resulting in 0.98 accuracy at optimal parameter}
\label{icml-historical}
\end{center}
\vskip -0.2in
\end{figure}
\begin{figure}[ht]
\vskip 0.2in
\begin{center}
\centerline{\includegraphics[width=\columnwidth]{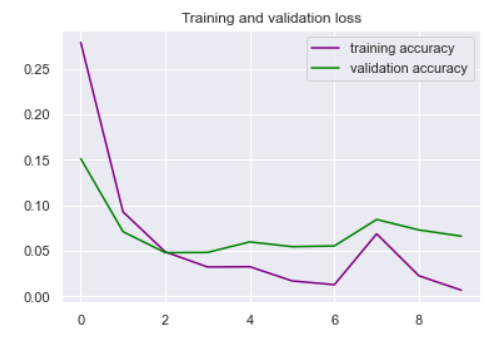}}
\caption{LSTM network resulting in 0.08 loss at optimal parameter.}
\label{icml-historical}
\end{center}
\vskip -0.2in
\end{figure}
The built LSTM network model is at the 4th layer of the framework where all processed output from each of the previous layers are merged and passed to the newly built LSTM model to make the final prediction.
\section{Framework Adaptability and Performance Evaluation}
To stretch and validate our multilayered adaptive framework for its effectiveness in the detection of phishing sites containing any of (Text, videos, and images), or a combination of any, or all of the 3. It is worth remembering that all existing AI or machine learning-based phishing detection techniques and frameworks can only detect text-based \cite{alsubaei2024enhancing}, \cite{jayaraj2024intrusion},\cite{jain2022survey},\cite{van2024applicability} or URL-based \cite{shukla2024http}, \cite{alsubaei2024enhancing},\cite{zhu2024pdhf},\cite{adebowale2023intelligent} phishing sites leading to their vulnerabilities to;\newline
-phishing sites with friendly URL \newline
-phishing site on hacked legitimate domain name server (DNS) \newline
-Image-only phishing site \newline
-video-only phishing site \newline
or, combination of any of them in any order. To validate both the effectiveness and adaptability of our proposed framework in overcoming such limitations, we created 4 categories of phishing sites and uploaded them to a secure server with a compromised DNS on a friendly URL; the first was a text-only phishing site, image-only phishing site, video-only phishing site, and a phishing site combining all the features.
We use friendly URLs, and hacked legitimate DNS on the text-only phishing site, so that they can evade detection at layer 1 until the 4th layer where it will be detected, while we created phishing sites containing each image-only, text-only, and a combination of both to test the adaptability of the framework to different scenarios of a phishing site.

In each scenario, we have 100\%  accuracy as the framework successfully adapts to each scenario and detects accordingly, thereby overcoming limitations associated with current approaches to phishing detection methods.

\section{Conclusion}
 In this research, we proposed a multi-layer adaptive framework that uses the computer vision capability of Optical Character Recognition (OCR) to read images on live phishing sites to text, and synthesize speech from uploaded deep-fake videos, while using Random Forest, and LSTM network, along with web scrapped text at various predictions layered of the framework to significantly improve the detection rate and performance of AI-based models for phishing detection. Considering the fact that existing AI-based phishing detection techniques, frameworks, and approaches can only detect text-based \cite{alsubaei2024enhancing}, \cite{jayaraj2024intrusion},\cite{jain2022survey},\cite{van2024applicability} or URL-based phishing \cite{shukla2024http}, \cite{alsubaei2024enhancing},\cite{zhu2024pdhf},\cite{adebowale2023intelligent} sites which leads to their vulnerability and inability to detect image-based, or video-based phishing sites, the proposed framework is able to overcome limitations in existing approaches, significantly improve phishing attack detection, and successfully detect complex phishing webpages with multi-dimentional deep-fake videos, images, and texts.

\section{Limitation and Future Research direction}
We used Mendeley and Kaggle phishing datasets which are URL-based and Text-based respectively. image-based and video-based phishing datasets are not publicly available because they are newly adopted forms of phishing websites to evade detection, we simulated them to get the data internally generated for this research especially with regard to deep-fake videos, hence getting publicly available image-based or Video-based phishing datasets will significantly help the research community in this.

The other research direction we will point at is the computational aspect during training. The proposed framework uses Random Forest and LSTM network at Layer 1 and Layer 4 respectively. The fact that Random Forest algorithm creates multiple trees each time to combine individual tree decisions for more accurate prediction leads to an increment in computation time, we have to set the random state to zero while changing the maximum depth for optimal hyper-parameter. Apart from the training computation time, there is also the server response time as the framework web scrapped the content behind the scenes thereby protecting the server against potential drive-by attacks. Hence, reducing the server response and computational to fraction of a second is an area open to future research in this domain. \newline

 It is also worth noting that our artifacts which consist of source code, dataset, images, videos, and audio files for this research had been uploaded to a public GitHub repository for reproducibility of our research. Artifact can be found on GitHub at; \newline 
\textbf{ \href{https://github.com/IGETOSIN1/Deep-Learning-Based-Speech-and-Vision-Synthesis-to-Improve-Phishing-Attack-Detection }{Deep Learning-Based Speech and Vision Synthesis to Improve Phishing Attack Detection through a Multi-layer Adaptive Framework} and also at Code Ocean computational research platform,} with the exception of the internally generated deep-fake video and audio data files for privacy.

\bibliographystyle{plain}
\bibliography{references.bib}

\end{document}